\documentclass[pdflatex,sn-mathphys-num]{sn-jnl}

\usepackage{array}
\usepackage{graphicx}%
\usepackage{multirow}%
\usepackage{amsmath,amssymb,amsfonts}%
\usepackage{amsthm}%
\usepackage{mathrsfs}%
\usepackage[title]{appendix}%
\usepackage{xcolor}%
\usepackage{textcomp}%
\usepackage{manyfoot}%
\usepackage{booktabs}%
\usepackage{algorithm}%
\usepackage{algorithmicx}%
\usepackage{algpseudocode}%
\usepackage{listings}%
\usepackage{comment}

\theoremstyle{thmstyleone}%
\theoremstyle{thmstyletwo}%

\theoremstyle{thmstylethree}%

\begin{document}

\title[A magnetically levitated conducting rotor with ultra-low rotational damping circumventing eddy damping]{A magnetically levitated conducting  rotor with ultra-low rotational damping circumventing eddy loss}


\author*[1]{\fnm{Daehee} \sur{Kim}}\email{daehee.kim@oist.jp}
\author* [1]{\fnm{Shilu} \sur{Tian}}\email{shilu.tian@oist.jp}

\author[1]{\fnm{Breno} \sur{Calderoni}}
\author[1]{\fnm{Cristina Sastre} \sur{Jachimska}}
\author[2]{\fnm{James} \sur{Downes}}
\author*[1]{\fnm{Jason} \sur{Twamley}}\email{jason.twamley@oist.jp}

\affil[1]{\orgdiv{Quantum Machines Unit}, \orgname{Okinawa Institute of Science and Technology Graduate University}, \orgaddress{\city{Onna}, \postcode{904-0495}, \state{Okinawa}, \country{Japan}}}

\affil[2]{\orgdiv{School of Mathematical and Physical Sciences}, \orgname{Macquarie University}, \state{NSW} \postcode{2109}, \country{Australia}}

\abstract{
Levitation of macroscopic objects in a vacuum is key towards the development of {high-precision} inertial sensors and pressure sensors, as well as towards the fundamental studies of quantum mechanics and its relation to gravity. Diamagnetic levitation offers a passive method at room temperature to isolate macroscopic objects in vacuum environments, yet eddy current damping remains a critical limitation for electrically conductive materials. We show that there are situations where the motion of conductors in magnetic fields does not, in principle, produce eddy damping, and demonstrate an electrically conducting rotor diamagnetically levitated in an axially symmetric magnetic field in a high vacuum. Experimental measurements and finite-element simulations reveal gas collision damping as the dominant loss mechanism at high pressures, while residual eddy damping, which arises from symmetry-breaking factors such as platform tilt or material imperfections, dominates at low pressures. 
The conclusion is supported by an analytic proof and an analytic example of zero steady current density for a rotating conductor in a magnetic field.
This demonstrates a macroscopic levitated rotor with extremely low rotational damping and paves the way to fully suppress rotor damping, enabling ultra-low-loss rotors for gyroscopes, pressure sensing, and fundamental physics tests.
}




\maketitle

\section*{Introduction}\label{Section:Introduction}
Rotors, freely rotating objects, require some form of trapping or levitation to pivot without restoring torque or libration. 
They are of significant interest both for practical applications and fundamental research.
They can be used as precision sensors either for gyroscopes \cite{Geim2001DetectionGyroscope, Zhang2023HighlyNanodiamond, Zeng2024OpticallyRotor}, measuring gas pressures \cite{Beams1962SpinningGauge, Fremerey1982SpinningGauges}, or for fundamental purposes such as probing vacuum friction \cite{Manjavacas2010ThermalParticles, Silveirinha2014TheoryFriction}, quantum revivals due to nonlinear dynamics \cite{Stickler2018ProbingNanorotors}, or collapse models  \cite{Carlesso2018Non-interferometricOptomechanics}. 
Recent research has primarily focused on nanoscopic rotors, with researchers demonstrating  GHz rotation speeds using optical trapping \cite{Jin20216Vacuum, Ahn2018, Reimann2018GHzVacuum}. 
Researchers have also demonstrated electrically trapped nanoscopic rotors and controlled their rotation speed via phased voltages on nearby electrodes \cite{Rider2019ElectricallyRotors}. 
However, optical and electrical trapping can only support objects whose size is limited to tens of microns, and the only force able to support large macroscopic rotors on the scale of mm-cm in a vacuum is magnetic forces. 
Magnetically levitated rotors have widespread use in industry for magnetic bearings, motors, and sensors. 
However, due to Earnshaw's theorem, if these systems are made of materials with positive magnetic susceptibility (paramagnetic, superparamagnetic, ferromagnetic), they require active feedback control for stable levitation \cite{Simon2001DiamagneticallyLevitation}. 
Using active stabilization, researchers have magnetically levitated a steel spherical rotor ($\sim 0.5\:{\rm mm}$ diameter) and spun it up to MHz rotational speeds using phased electromagnets \cite{Schuck2018UltrafastSpheres}.

Alternatively, diamagnetic levitation, which requires no active control or power, could be used to levitate the rotor. However, most room-temperature diamagnetic materials are electrical conductors and will typically experience eddy damping. 
Recently, researchers have found ways to reduce the eddy damping in diamagnetically levitated systems, either by modifying the geometry of the object itself to interrupt the eddy current flow \cite{Romagnoli2023ControllingPlate, Xie2023SuppressingGeometry} or by making the object out of fine diamagnetic powders \cite{Chen2022DiamagneticResonators, Tian2024FeedbackPlate}, limiting the eddy current flow to the individual powder grains. 
Developing diamagnetically levitated rotors using such techniques is possible, but will significantly lower the levitation strength. 

To retain maximal levitation lift, we instead consider the levitation of a bulk solid diamagnetic electrical conductor, such as pyrolytic graphite, and ask: {\em are there arrangements of magnetic fields and levitated diamagnetic objects, such that the dynamics of the levitated objects evoke {no eddy damping}, particularly for rotational motion? }

It is well known that eddy damping can be affected by the physical shape of the conductor, but this can be more nuanced. 
Consider a square slab of graphite diamagnetically levitated above a checkerboard array of magnets as in \cite{Tian2024FeedbackPlate}. Although not a rotor (as all degrees of freedom are trapped), it exhibits eddy damping for both translational and librational dynamics. 
Reshaping this graphite into a disk will ideally remove any angular dependence on the motion. 
However, one can clearly observe that eddy damping is still present, rapidly reducing the rotational oscillations (see Supplementary movies {1, 2}). If we look at a small region of this levitated diamagnetic conducting disk, as the disk rotates, this region will experience a change in magnetic flux as the checkerboard's magnetic fields are not axially symmetric, and this changing flux will produce eddy currents. 
If we instead consider a conducting disk that is rotating in a magnetic field that is either homogeneous (and transverse to the plane of the disk) or has axial symmetry about the axis of rotation of the disk, then each small region of the disk will experience no change in magnetic flux as the disk rotates. Thus, no eddy currents should be generated due to this rotational motion, and only gas damping will be present, which, at low pressures, can be exceedingly small.

Interestingly, the literature has some disagreement on the correctness of this hypothesis. There have been a number of theoretical studies that predict eddy damping torques to vanish in the presence of a magnetic field co-axial to the rotation axis for a rotating conducting disk or sphere
\cite{Smythe1942OnDisk, Schieber1974BrakingField, Chen2009EvaluationMicrosystems, Redinz2015FaradaysAnalysis, Li2025StudyGauge}. We are aware of only a few related experimental studies.  In 1966, Waldron reported low rotational damping of a diamagnetically levitated ring bearing \cite{Waldron1966DiamagneticGraphite}. 
The development of the spinning rotor vacuum gauge, which is useful for measuring gas pressures in ultra-high vacuums, led Fremerey in 1971 and 1972 \cite{Fremerey1971ApparatusSuspension, Fremerey1972ResidualSpheres}, to develop feedback-stabilized magnetic levitation of small steel spheres which exhibited low rotational damping. 
Researchers have reported the diamagnetic stabilized levitation of a cylindrical permanent magnet and its spin-down in high vacuum over hours \cite{Cansiz2004StableBearings}, while others investigated diamagnetically levitated graphite disks using coaxially symmetric magnetic fields in air, which appeared to display rotational eddy damping \cite{Nishiwaki2023MaterialsCamera}, which is in conflict with theoretical expectations. 
These experimental works did not investigate the ultimate limits of rotational damping.
In Supplementary Note 2 and Supplementary Table 2, we compare the minimum damping rates for different levitated rotational systems.

In our study, we investigate whether one can reach ultra-low rotor loss in a setup where a conducting diamagnetic rotor is levitated above a cylindrical magnet array in a controlled high-vacuum environment.  
We measure the rotational damping rate via the spin-down dynamics of the disk at various pressures. 
We perform finite element simulations [COMSOL] to obtain estimates for this rotor damping rate at high and intermediate pressures and find good agreement with the measurements. 
At the lowest pressures, we find that the rotor damping rate appears to be again dominated by small eddy current damping, which depends, in part, on the inclination of the setup to the horizontal. 
When the setup experiences a small tilt, the levitated rotor is shifted to rotate off-axis by gravity, with subsequent small eddy damping loss. 
To confirm that this loss vanishes in the ideal limit, we show an analytic proof and give an analytic model example to show explicitly that a uniformly rotating conductor carries zero steady-state current. 
This example demonstrates that for a perfectly axisymmetric array, eddy current damping is identically zero.
Demonstrating a diamagnetic rotor with extremely low rotational loss can pave the way towards the engineering of rotational quantum superpositions \cite{Stickler2018ProbingNanorotors}, ultra-precise gyroscopes \cite{Dohogne1982MagneticallyGyroscope, Geim2001DetectionGyroscope, Zhang2023HighlyNanodiamond,Zeng2024OpticallyRotor}, and pressure sensors \cite{Fremerey1982SpinningGauges}, {and the investigation of} vacuum friction \cite{Zhao2012RotationalFriction, Stickler2018RotationalRotors, Pan2019MagneticallyFriction, Khosravi2024GiantNanospheres}, {and} quantum decoherence of rotational systems \cite{Glikin2025ProbingRotor}.

\section*{Results}\label{Section:Results}

\subsection*{Detection and fabrication of diamagnetically levitated disk}\label{Subsection:Setup}

We magnetically levitate a diamagnetic, conductive pyrolytic graphite (PG) disk within an axisymmetric magnetic trap, as illustrated in Fig. \ref{fig:experimentalSetup}.
The trap consists of five layers of ring-shaped NdFeB magnets ($\mathrm{OD}\, 19\times \mathrm{ID}\,8.1\times \mathrm{H}\, 4$ mm; N40) and two layers of cylindrical NdFeB magnets ($\mathrm{D}\, 8\times \mathrm{H}\, 10$ mm; N52), arranged with alternating magnetization for radial trapping.
Due to its strong diamagnetic properties at room temperature, the PG disk achieves stable macroscopic levitation using commercial permanent magnets.
The disk (diameter $D= 10.02$ mm, height $H= 1.12$ mm) was precisely fabricated from a PG plate via Computer Numerical Control (CNC) machining, followed by diamond polishing to ensure high circularity.
Further fabrication details are provided in the Methods{: PG disk fabrication process}.

The rotation of the disk was measured using an event-based camera (EBC), which detects local brightness changes.
To facilitate tracking, a small white ink dot was applied to the top surface of the disk, as shown in the overhead view of Fig. \ref{fig:experimentalSetup}a. 
The EBC extracted the dot's center coordinates using a built-in motion-tracking algorithm, enabling the determination of the disk’s kinematics.
Figure \ref{fig:experimentalSetup}b shows an example of twenty tracked dot positions over $100 \:\rm ms$.
The disk’s top surface was viewed via a 45$^\circ$-angled flat mirror assisted by two convex lenses, L1 and L2, with corresponding focal lengths of $f_1=3$ mm and $f_2=12$ mm, forming a Keplerian telescope for magnification.

To analyze the rotational dynamics of the levitated disk, we define the rotational damping rate $\it{\gamma}$ [Hz], which characterizes the exponential decay of the angular velocity $\omega (t)$:
\begin{equation}
\label{eq:DefineGamma}
    \omega(t)=\omega_0 e^{-\gamma t},
    \nonumber
\end{equation}
where $\omega_0=\omega(t=0)$ is the initial angular velocity.
The exponentially decaying dynamics arise when the damping torque is viscous, meaning it is linearly proportional to the angular velocity:
\begin{equation}
    \mathcal{T}=-{\it{\Gamma}} \omega, \nonumber
\end{equation}
where ${{\it{\Gamma}}}=\gamma I$ [$\rm N\cdot m\cdot s$], is the rotational damping coefficient, and $I=\frac{1}{2}MR^2$ is the moment of inertia of a uniform disk (mass $M$, radius $R$), about an axis of rotation passing through the center of the disk,  perpendicular to the plane of the disk.
The negative sign indicates that the torque acts to dampen the rotation.
In our experiment, the assumption of viscous damping holds accurately, as evidenced by the linear relationship:
\begin{equation}
    \frac{d}{dt}\ln \omega(t)\propto -{\rm const}. \nonumber
\end{equation}

\subsection*{Pressure dependence of rotational damping}
Gas friction damping is one of the major motional damping sources caused by the collision between the gas molecules and the object, and is dependent on the gas pressure. 
To distinguish gas friction damping from other rotational damping mechanisms, we first characterize the pressure dependence of the disk's rotational damping rate $\gamma$.

We determine the rotational damping rate $\gamma$ by monitoring the decay of the disk's angular velocity $\omega$ using the EBC. 
Rotation was induced by lateral vibrational excitations of the setup along the $x-$ and $y-$ axes. This produced both rotational and translational motions, though the latter damped out rapidly. 
The angular velocity $\omega(t)$ was extracted from the disk's orientation $\phi(t)$, detected by the EBC.
In Figs. \ref{fig:gas_damping_result}a, b, we plot the decay curve of $\omega(t)$ at two typical pressures, $1.0\times10^{5} \:\rm Pa$ and $5.3\times10^{-5} \:\rm Pa$, respectively.
The angular speed exhibits clear exponential decay, and $\gamma$ is obtained from the slope of the linear fit of $\rm ln(\omega)$. 
The fit to exponential decay is very clear and over a large time span. 
The obtained damping rates at different pressures are plotted in Fig. \ref{fig:gas_damping_result}c. 
At high pressure, $\gamma$ remains nearly constant, while in the intermediate pressure range, it scales linearly with pressure.
At low pressure, $\gamma$ reaches a plateau.

The interaction mechanism between gas molecules and an object strongly depends on the ratio between the gas molecule's mean free path $\it{\lambda}$ and the 
size of the object of interest $H$, which is known as the Knudsen number, $\textrm{Kn}={\lambda}/{H}$.
At high pressure ($P>10^{3} \:\rm {Pa}$), where $\textrm{Kn}<0.01$, gas behaves as a continuous fluid and follows the principles of classical fluid dynamics, a regime known as continuum flow.
As pressure decreases ($10^{0}\:\textrm{Pa}<{P}<10^{3} \:\rm Pa$), the system enters a transition regime where the gas flow gradually shifts from continuum behavior to free molecular flow.
At low pressure ($P<10^{0} \:\rm Pa$), where $\textrm{Kn}>10$, intermolecular collisions become negligible, and gas molecules move ballistically, defining the free molecular flow regime.

We developed a FEM model to simulate the rotational gas friction damping of a levitated disk in the continuum flow regime.
Under these conditions, airflow can be well-behaved and laminar, chaotic and turbulent, or in a transitional state between these extremes, depending largely on flow velocity.
Our experiments apply to the laminar flow regime as the angular velocity of the rotating disk is very low. 
COMSOL's laminar flow module was used to model this behavior.
Figure \ref{fig:comsol_gas_friction}a depicts a cross-sectional view of the simulated laminar flow generated by a rotating disk at $\omega=2{\pi}\times2$ {$\rm{rad}\cdot s^{-1}$} under ambient conditions.
At $\textit{z}={0}$ mm, a no-slip boundary condition is applied to simulate the presence of the top surface of the magnetic array.
The color density shows the velocity magnitude of the flow, whereas the streamline shows the direction of the axial flow.
By integrating the airflow-induced shear stress over the disk's surface, we obtain the total rotational gas friction torque $|\mathcal{T}(\omega)|=
{\it{\Gamma}} \omega=\gamma I\omega$, as a function of $\omega$ at a specific pressure, as shown in Fig. \ref{fig:comsol_gas_friction}b.
The slope of the linear fit corresponds to ${\it{\Gamma}}=\gamma I$.
The simulated damping rate $\gamma$ at different pressures is plotted as blue circles as shown in Fig. \ref{fig:comsol_gas_friction}c.
The simulation agrees well with the experimental data in the continuum flow regime at high pressures $P>10^2 \:\rm Pa$ (unshaded region). 
In this regime, the rotational damping rate changes slightly with pressure.
Refer to the Methods for further details on the COMSOL modeling.
In the free molecular flow regime, the damping coefficient of the rotational motion of a disk is linearly proportional to the pressure $P$ \cite{Cavalleri2010GasReservoir},
\begin{equation}
\label{eq:cylinder_rot_gas_damping_coeff}
    {\it{\Gamma}}_{\mathrm{fm}}(P) = \sigma R^4 
 \sqrt{\frac{\pi m_0}{2 k_{{\textrm{B}}} T}}\left(1+\frac{2H}{R}\right) P,
 \nonumber
\end{equation}
where $\sigma$ is a dimensionless accommodation factor, 
$m_0$ is the mass of the gas molecule, $T$ is the temperature of the surrounding environment, and  $R$ and $H$ are the radius and thickness of the disk, respectively. 
The accommodation factor $\sigma$ describes the surface effects, and we use $\sigma=1$ for our polished-surface sample. 
We analytically obtain the rotational damping rate in the free molecular flow regime as $\gamma_{\mathrm{fm}}(P) = {\it{\Gamma}}_{\mathrm{fm}}(P)/I$.
As shown in Fig. \ref{fig:gas_damping_result}c, the analytical prediction matches well with the experimental data in the range of $10^{-1}\:\textrm{Pa} <{\it{P}}<5\times10^0\:\rm Pa$, which is the free molecule flow regime, implying negligible surface effects and squeezed-film-like effects in our setup.
However, the experimental data reach a plateau at lower pressure, suggesting that other forms of damping dominate the gas molecule damping at very low pressure.

\subsection*{Rotational eddy damping via tilt-induced symmetry breaking}
When the axially symmetric levitation setup is placed on a platform that has a slight inclination to the horizontal, the center of mass (COM) of the levitated disk is displaced laterally due to gravity, leading to the breaking of axial symmetry. In such a situation, we expect there to be small eddy damping associated with the broken axial symmetry of the disk's rotation.
To explain the plateau in the rotational damping in the low-pressure region (see  Fig. \ref{fig:gas_damping_result}c), we study the dependence of the rotational damping rate on small tilting angles.

To investigate the effect of the small tilt, we keep the pressure constant $P\sim 4.85(\pm 0.25)\times 10^{-5}$ Pa, which is within the low-pressure plateau regime.
As depicted in Fig. \ref{fig:experimentalSetup}, the vacuum chamber containing the experimental setup sits on a thick aluminum plate, which itself sits on an optical table. 
We controlled the tilt of the setup in 2D, characterized by small tilt angles about the $x-,\; y-$ axes,
$\theta_x$ and $\theta_y$, via adjustments of the nitrogen flow level of the optical table legs (coarse tuning) and using screw jacks (fine tuning). 
We then measured the rotational damping rate $\gamma(\theta_x,\theta_y)$,  at various two-dimensional inclinations of the platform {$(\theta_x^i, \theta_y^j)$}, as shown in Fig. \ref{fig:TiltVariation}c.
Red circles and blue squares indicate a separate tilt sequence {$(\theta_x^i, \theta_y^j)$}, and their measured values of $\gamma$, and a fitted 2D contour plot.
We observed that the minimum damping was evidenced at a non-zero 2D tilt  $\theta_x^{\rm{min}},\theta_y^{\rm{min}} \neq (0,0)$, which corresponds to the orientation at which gravity is perpendicular to the top of the magnet array inside the chamber (i.e., the “true absolute level”).
This offset is due to a slight misalignment of the inclination of the platform inside the vacuum chamber (further discussed in the Methods{: Inclination angle measurement}).
From this minimum value, $\gamma$ shows a steep increase by an order of magnitude within ${\it{\Delta}}\theta \sim 0.5^\circ$, showing a clear tilt angle dependence of $\gamma$.

We fitted the two-dimensional damping rate $\gamma(\theta_x,\theta_y)$ function to a one-dimensional damping rate $\tilde{\gamma}({\it{\Delta}} \theta)$, assuming that the damping rate is axially symmetric about the minimum   ($\theta_x^{\rm{min}},\theta_y^{\rm{min}}$), and is only a function of the tilt angle away from this minimum location. We plot $\tilde{\gamma}({\it{\Delta}} \theta)$, in Fig. \ref{fig:TiltVariation}d, and observe that all the data fits well onto a smooth curve, justifying the assumption of axial symmetry in tilt for $\gamma$.
We compare this experimental data with simulated results obtained via FEM, represented by the black dashed curve.
The FEM simulation models the eddy damping force due to the off-axis rotation of the disk. To perform this simulation, one requires estimates for the values of various material properties, such as the magnetic susceptibilities of the graphite.
The diamagnetic susceptibilities in the horizontal and vertical directions in the graphite are estimated to match the measured horizontal and vertical oscillation frequencies of the disk and the measured levitation height, given the strength of the magnet arrays.
Using the force diagram depicted in Fig. \ref{fig:TiltVariation}b, we obtain a relationship between the tilt angle of the platform, ${\it{\Delta}}\theta$, and the displacement of the center of mass  of the levitated disk from the axis of axial symmetry, $d$,
\begin{equation}
    d({\it{\Delta}}\theta)=g
    {\it{\Delta}}\theta /\omega_{{\textrm{L}}}^2
\label{eq:Tilt_LateralDisplacement_Relation}
\end{equation}
where $g$ is the gravitational acceleration.
In the above relation, a small-angle approximation is also used, and its experimental validation is discussed in the Methods:{Tilt-horizontal displacement relation}.
As we work in the small-angle regime, the relative tilting of the levitated disk with respect to the top surface of the magnetic array is also neglected.
The FEM simulation of the resulting off-axis rotational eddy-damping also requires values for the in-plane $S_\parallel$ and perpendicular $S_\perp$ electrical conductance of the disk. 
We find that $\gamma$ is insensitive to $S_\perp$, and we are thus left with one fitting parameter $S_\parallel$, whose fitted value is shown in Table \ref{table:simulation_parameters}.
The simulated damping matches well with experimental results for ${\it{\Delta}} \theta > 0.1^\circ$.

On the other hand, the 3D FEM simulations become unreliable below ${\it{\Delta}}\theta<0.1^\circ$.
In Fig. \ref{fig:TiltVariation}e, we plot the simulated damping $\gamma_{\rm{sim}}(d)$ as black squares and its zero-displacement limit $\gamma_{\rm{sim}}(0)$ as a dashed line.
Although perfect axial symmetry demands $\gamma(d\rightarrow 0)= 0$, the nonzero damping here stems from mesh-induced asymmetry and numerical error (see Supplementary Material).
To bypass these artifacts, we fit the region $d>0.05\:\rm{mm}$ to a power law,
\begin{equation}
    \gamma_{\mathrm{fit}}(d)=c_1d^{c_2},\quad c_1=6.20\times 10^{4},\quad c_2=1.91,
    \label{eq:power_law_fit}
\end{equation}
(shown as the gray line).
The near-perfect power-law dependence strongly supports $\gamma(d=0)=0$ under ideal symmetry.

When the setup gets close to the axially symmetric conditions, the damping rate drops dramatically, and the errors in the FEM simulations dominate, leading to a non-zero damping plateau (as shown in Fig. \ref{fig:TiltVariation}d and e). 
We then build a 2D axisymmetric FEM model to simulate the eddy damping, which will eliminate the error caused by the asymmetric mesh. 
The results show a lower damping rate of $\gamma \sim 4\times 10^{-6} \:\rm Hz$, which is still far from zero, which is due to numerical errors in the FEM simulation. (See Supplementary {Note 1})
To confirm this, we next present analytical proofs and a demonstration that shows that in the case of perfect axial symmetry, $\mathbf{J}=0$.

\subsection*{Analytical demonstration of vanishing current in an ideal symmetry}
We now analytically prove that the steady-state current density $\mathbf{J}$ inside an axially symmetric rigidly rotating conductor, rotating co-axially in an axially symmetric magnetic field, is zero for the entire volume.
We consider an axisymmetric field 
\begin{equation}
    \mathbf{B}(r,z)=B_r(r,z)\,\hat r + B_z(r,z)\,\hat z,
    \quad 
    \nabla
    \cdot\mathbf{B} = \tfrac1r\partial_r(rB_r)+\partial_zB_z=0.\nonumber
\end{equation}
For a conductor with rigid rotation, $\mathbf{v}=\omega r\,\hat\phi$, the current density expression is:
\begin{equation}
    \mathbf{J}=\sigma\bigl(-\nabla V + \mathbf{v}\times\mathbf{B}\bigr),
    \quad
    \mathbf{v}\times\mathbf{B} =v_\phi B_z\,\hat r - v_\phi B_r\,\hat z.\nonumber
\end{equation}
Using \(\nabla\times(f\mathbf{A})=\nabla f\times\mathbf{A}+f\nabla\times\mathbf{A}\), we get
\begin{equation}
\begin{aligned}
\nabla\times\mathbf{J}
&=\nabla\times\bigl(-\sigma\nabla V + \sigma(\mathbf{v}\times\mathbf{B})\bigr)\\
&= -(\nabla\sigma\times\nabla V)
  -\sigma(\nabla\times\nabla V)
  +\nabla\sigma\times(\mathbf{v}\times\mathbf{B})
  +\sigma\,\nabla\times(\mathbf{v}\times\mathbf{B})\\
&= \nabla\sigma\times\bigl[-\nabla V + (\mathbf{v}\times\mathbf{B})\bigr]
  -\sigma(\nabla\times\nabla V)
  +\sigma\,\nabla\times(\mathbf{v}\times\mathbf{B}).
  \nonumber
\end{aligned}
\end{equation}
Enforcing \(\nabla\sigma=0\) for a uniform conductivity and noting \(\nabla\times\nabla V=0\), only the last term remains:
\begin{equation}
    \nabla\times\mathbf{J} = \sigma\,\nabla\times(\mathbf{v}\times\mathbf{B}).
    \nonumber
\end{equation}
But
\begin{equation}
    \nabla\times(\mathbf{v}\times\mathbf{B})
= {\omega r} \Bigl(\partial_zB_z + \tfrac1r\partial_r(rB_r)\Bigr)
= {\omega r}\,\nabla\!\cdot\!\mathbf{B}
= 0.\nonumber
\end{equation}
Hence \(\nabla\times\mathbf{J}=0\) for a rotating conductor in an axially symmetric magnetic field. This now implies that there cannot be any closed current loops within the conductor, e.g., eddy currents. However, we can go further and show that $\mathbf{J}=0$.

Since $\nabla\times\mathbf{J}=0$, there exists a scalar potential ${\it{\Phi}}$ such that $\mathbf{J}=\nabla{\it{\Phi}}$.
The steady-state continuity and boundary conditions for the current satisfy
\begin{equation}
    \nabla\cdot\mathbf{J}=0,\quad \mathbf{J}\cdot\hat{n}|_{S}=0 
    \quad\Rightarrow \quad
    \nabla^2{\it{\Phi}}=0,\quad \nabla{\it{\Phi}}\cdot\hat{n}|_{S}=0.
    \nonumber
\end{equation}
Using the above conditions and Green's first identity, we show that
\begin{equation}
    \int_V \left|\nabla {\it{\Phi}}\right|^2\;dV
    =\int_S {\it{\Phi}}\left(\nabla{\it{\Phi}} \cdot \hat{n} \right)\:dS -\int_V {\it{\Phi}}\nabla^2{\it{\Phi}}\:dV
    =0 \nonumber.
\end{equation}
This implies that $\nabla{\it{\Phi}}=0$ everywhere in $V$, and therefore,
\begin{equation}
    \mathbf{J}=\nabla{\it{\Phi}}=0\quad \text{throughout the conductor}.\nonumber
\end{equation}

Following the proof, we demonstrate an explicit calculation of an example in which a rotating conductor carries no steady‐state current.
Guided by the oscillatory radial components, we define an axisymmetric magnetic field that is divergence‐free by construction:
\begin{equation}
    B_r(r)=\beta \sin{\alpha r},\quad
    B_z(r,z)=-\beta (z+z_0)\left( \frac{\sin{(\alpha r)}}{r}+\alpha \cos{(\alpha r)}\right).
    \label{eq:analytical_B_field_example}
\end{equation}
As shown in Fig. \ref{fig:B_field_comparison_analytical}, the vector components of the field (with $\alpha=\pi/6,\;\beta=0.1,\;z_0=-4$) closely replicate those of the magnetic field generated from an axially symmetric magnetic array.

Assuming for this example isotropic conductivity $\sigma$ and rigid rotation about the axis, $\mathbf{v}=\omega r \hat{\phi}$, we calculate that 
\begin{equation}
    \mathbf{v}\times\mathbf{B}=-\omega z(\sin{(\alpha r)}+\alpha r \cos{(\alpha r)})\hat{r}-\omega z \sin{(\alpha r)}\hat{z}\nonumber.
\end{equation}
Using this, we obtain the continuity equation to be
\begin{equation}
    \nabla^2V= \nabla\cdot\left(\mathbf{v}\times\mathbf{B}\right)=-\omega z\left( \frac{\sin{(\alpha r)}}{r}+3\alpha \cos{(\alpha r)}-\alpha^2 r \sin{(\alpha r})\right).\nonumber
\end{equation}
We then seek $V(r,z)$ satisfying this Poisson equation and $J_r(r=0,R)=J_z(z=\pm H/2)=0$.
A direct check shows that the ansatz
\begin{equation}
    V(r,z)=-\omega r z \sin{(\alpha r)}\nonumber
\end{equation}
both solves $\nabla^2V= \nabla \cdot \left (\mathbf{v} \times \mathbf{B} \right)$ and enforces $\mathbf{J}\cdot\hat{n}=0$ on all boundaries.
Substituting into the current density expression, we obtain that
\begin{equation}
    \mathbf{J}(r,z)=0\text{ for all }0\le r\le R\text{ and }-\frac{H}{2}\le z \le \frac{H}{2}.\nonumber
\end{equation}
Thus, even in the presence of rotation and a nontrivial magnetic geometry, no bulk currents circulate. 
This analytic result validates our FEM simulations and confirms that there is no eddy damping for a conductive disk rotating with a constant angular velocity in an ideally axially symmetric magnetic field.
{We thank the reviewer for bringing references \cite{PLorrain1990,PLorrain1993,VanBladel1984}  to our attention, which also include a proof.}

\subsection*{Possible origin of residual damping}
While ideal axial symmetry predicts zero eddy damping, our experiments reveal a residual rate of $\gamma_{\rm{min}}=5.5\times 10^{-5}\:\rm{Hz}$ equivalent to a static symmetry-breaking offset of $d\approx 18\:\mu \rm{m}$ (Eq. \ref{eq:power_law_fit}).
Side-view video shows the slight wobbling of the disk, which in fact reflects the disk's center of mass tracing a small circle around the trap axis at a fixed offset, equivalent to a static tilt that generates eddy-current damping  (Supplementary movie 3).
Such an offset may arise from slight geometric or material asymmetries in either the pyrolytic graphite disk or the magnet assembly. 
For example, lateral variations in conductivity or diamagnetic susceptibility may be present due to misalignment between the disk’s geometric axis and the graphite C-axis.
Indeed, when the disk is levitated on the checkerboard magnet array, the disk has a preferential orientation with librational confinement, signaling an internal asymmetry (see Supplementary movie 1).
{Other possible sources of damping, such as surface-localized currents from the skin effect at high angular velocities in an inhomogeneous electromagnetic environment, were examined in Supplementary Note 3 and found to be negligible in our setup.}

Dynamical symmetry breaking from thermal fluctuations or environmental vibrations can displace the disk off-center, inducing eddy-current damping even with perfect static tilt control.
To quantify the damping rate arising from thermal fluctuations at ambient temperature, we estimate the root-mean-square displacement via the equipartition theorem, $x_{\rm{rms}}=\sqrt{k_{{\textrm{B}}} T/mw^2}\approx 1.2\times 10^{-10}\;\textrm{m}$ for $T=300\:\textrm{K},\:{\it{m}}=191\;\textrm{mg}$, and $\omega/2\pi= 6\;\textrm{Hz}$.
Substitution into Eq. \ref{eq:power_law_fit} yields a thermal noise-induced damping rate of $\sim 7\times 10^{-15}\:\rm{Hz}$, some ten orders of magnitude below our measured $\gamma_{\rm{min}}$.
Moreover, gyroscopic stabilization at higher angular velocity further suppresses any residual eddy currents due to transverse motion.
We therefore conclude that static imperfections dominate the observed residual damping.

\section*{Discussion}\label{Section:Discussion}
We show a diamagnetically levitated pyrolytic graphite rotor with an extremely low rotational damping rate at a high vacuum. 
Through our study, we show that a perfect disk-shaped graphite rotor, although an electrical conductor, experiences no eddy damping when it rotates in an axisymmetric magnetic field.
The results show that the gas collision damping is strongly suppressed at low pressure.
However, we experimentally observe that small eddy damping is present, and even dominates at low pressure, which we attribute to some breakage of the axial symmetry of the setup.
Laser cutting \cite{Romagnoli2023ControllingPlate} or electrical discharge machining (EDM) \cite{Su2017ExplorationRotor} may be used to fabricate a sample with a better geometric symmetry.
To minimize material inhomogeneities in the disk, diamagnetic material with high crystalline quality (e.g., bismuth) may be used.
With the application of high-quality diamagnetic disks and magnets and robust rotation actuation methods \cite{Xu2019PassiveMotor, Sugimoto2023NovelRotor}, this setup has the potential to reach extremely low rotational damping with precise tilt angle control.
A rotational eddy damping rate of $\gamma\sim 10^{-11}\, \textrm{Hz}$ can be expected from extrapolation of Eq. \ref{eq:power_law_fit}, if the tilt of the setup is controlled within $\theta\sim\mu \mathrm{rad}$.
The nearly free-rotating disk has many potential applications, including ultraprecise sensing for inclination, acceleration, pressure, testing vacuum frictions, and engineering rotational macroscopic superpositions.

\section*{Methods}
\begin{figure*}[tb]
\centering
\includegraphics[width=\textwidth]{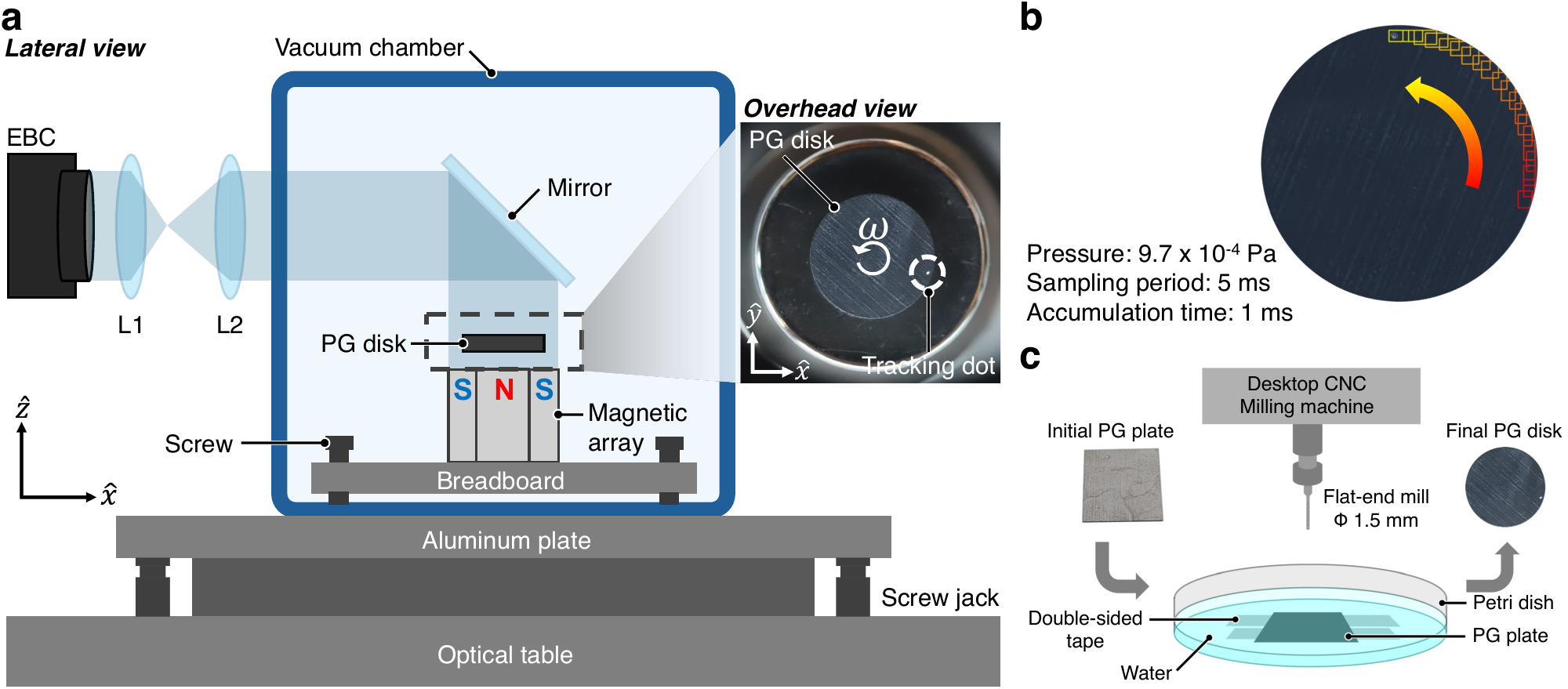}
\caption{
Diamagnetic levitation and rotation of a pyrolytic graphite (PG) disk in an axially symmetric magnetic trap.
\textbf{a}: Lateral view of the levitation setup.
An axially symmetric magnetic array, consisting of cylindrical and surrounding ring magnets arranged with alternating vertical magnetization, levitates a PG disk.
The magnets are fixed to a {breadboard}, whose initial inclination is adjusted to be as horizontal as possible using screws inside the vacuum chamber (reaching $P\sim$ 5 $\times 10^{-5}$ Pa).
The chamber itself is mounted on an aluminum plate atop an optical table, with its inclination adjusted away from the horizontal using the optical table and screw jacks. 
The top surface of the levitating disk is viewed via a flat mirror and two convex lenses (L1, L2), forming a Keplerian telescope for magnification, and its motion is tracked using an event-based camera (EBC).
The inset shows a photograph of the levitated PG disk, marked with a white ink dot for orientation.
\textbf{b}: 
EBC [Prophesee EVK-V3HD], detections of the tracking dot on the rotating PG disk. Twenty consecutive detections (squares) are shown for illustration.
\textbf{c}: Schematic of the PG disk fabrication. 
A square PG plate (3 cm sidelength) is roughly milled using a desktop CNC machine [SnapMaker].
It is secured in a Petri dish half-filled with water, which disperses cut material, cools the disk, and eliminates dust formation.
The disk is then precisely shaped and polished using a lapidary faceting machine [Ultratec V5], producing the final 10 mm diameter PG disk (right image).
}\label{fig:experimentalSetup}
\end{figure*}

\begin{figure*}[tb]
\centering
\includegraphics[width=\textwidth]{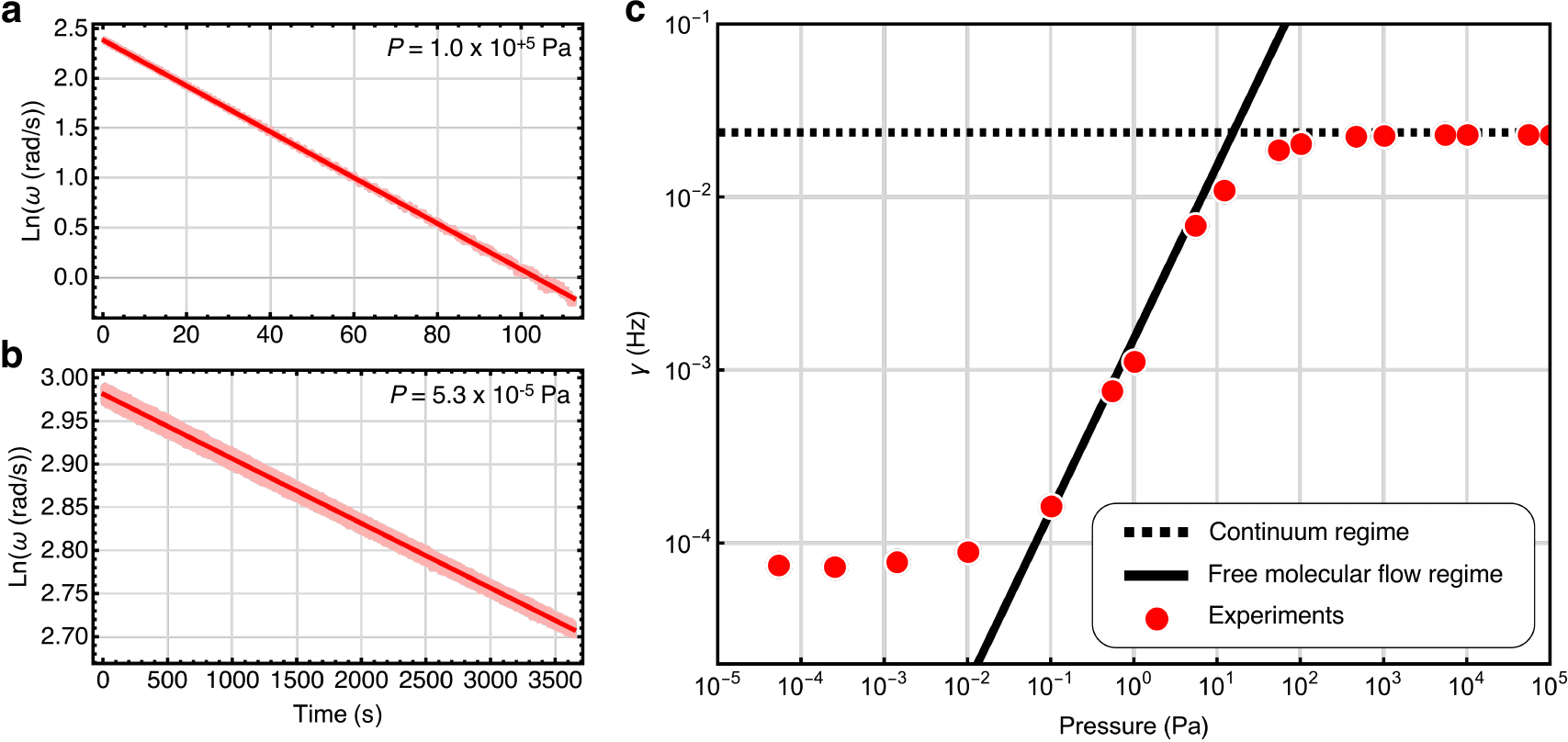}
\caption{
    Spin-down of an initially rotating levitated disk. 
    The disk is first spun up using an applied torque. 
    Once it reaches a sufficiently high angular velocity 
    {$\omega$}, the torque is removed, and its spin-down is monitored over time. 
    \textbf{(a, b)}: 
    Experimentally measured values of $\ln(\omega)$ as a function of time for high and low gas pressures $P$, along with best-fit lines whose slopes correspond to $\gamma(P)$.
    The data closely follow $\omega(t)= \omega_0 \exp(-\gamma t)$, indicating that angular velocity damping is well-approximated by viscous drag.
    \textbf{c}: Rotational damping rate $\gamma(P)$ as a function of gas pressure $P$ for a fixed, non-zero inclination of supporting platform.
    The horizontal dashed line represents the damping rate estimated via FEM-COMSOL simulation in the high-pressure regime, where the continuum assumption of fluid dynamics holds.
    The diagonal solid line represents the theoretical prediction of $\gamma(P)$ in the free molecular flow regime \cite{Cavalleri2010GasReservoir}.
    The experimentally determined values of $\gamma(P)$ from spin-down measurements are shown as red circles.
    {The associated error, obtained from the error in the linear fit, is smaller than the marker size.}
    For $P<10^{-1}$ Pa, a plateau in $\gamma(P)$ suggests a dominant damping mechanism beyond gas friction, likely eddy damping due to deviations from perfectly axial symmetry. 
    Potential sources include material imperfections in the disk or magnets, or a slight inclination of the setup, shifting the PG disk's center of mass off-axis. 
}
\label{fig:gas_damping_result}
\end{figure*}

{
\subsection*{PG disk fabrication process}
}
We fabricated the PG disk with high geometric axial symmetry using a two-step process: CNC milling followed by faceting, carefully accounting for the material’s fragility.
First, a desktop CNC milling machine [Snapmaker A250] with a 3-axis carver module and a 1.5 mm flat-end bit was used to mill the top and bottom surfaces of a rough PG plate, ensuring uniform thickness. 
The initial 50×50×3 mm PG plate was secured inside a Petri dish with double-sided tape. Water was added to cool the working area, disperse cut material, and prevent aerosolized graphite dust (Fig. \ref{fig:experimentalSetup}c). 
The edges of the Petri dish were clamped to maintain stability while milling both surfaces. 
To release the plate, ethanol was used to dissolve the adhesive.
Next, the smoothed PG plate was refined using a faceting machine [Ultratec V5 Digital] with a P3000 diamond lap to correct radial asymmetries and further smooth the surface with P2400 sanding paper. 
The sample was affixed to an 8 mm faceting dop using water-soluble wax. 
With the dop set at 90° to the $z-$axis, the disk’s radial edges were trimmed using free axial rotation. 
The dop was then adjusted to 0° to ensure a flat, axially symmetric surface. 
After machining, a heating pad melted the wax, allowing the sample to be flipped for final lapping and polishing on the opposite side.

{
\subsection*{Angular velocity measurement}
}
The disk’s rotation was recorded with a Prophesee EVK-V3HD event-based camera by tracking a white marker. 
Using the camera’s built-in algorithm, we extracted the marker’s $(x,y)$ coordinates and defined their mean position as the origin. 
We converted these Cartesian coordinates to a polar angle $\phi(t)$, then smoothed the raw $\phi(t)$ data with a Gaussian-weighted moving average (standard deviation $\sigma = 100\,$ms). 
From the smoothed angle, we calculated the instantaneous angular velocity $\omega = \dot\phi$ and evaluated $\ln\omega$. 
Finally, the damping rate was determined from the slope of a linear fit to $\ln\omega(t)$.

{
\subsection*{Inclination angle measurement}
}
The platform's inclination was measured using an inclination sensor \cite{Staacks2018AdvancedPhyphox}, providing two angular degrees of freedom ($\theta_x$, $\theta_y$).
Measurements were taken over 1-10 minutes, and the mean values were recorded with a typical standard deviation of $\sigma_\theta\approx0.05^\circ$.
Before use, the sensor's systematic internal angular offsets were characterized by measuring a fixed inclination $\theta_x\sim 5^\circ$ and then rotating the sensor by $\pi$, which should yield $\theta_x=\pm 5^\circ$ in the absence of systematics. 
Measured offsets were $\theta_{x,\textrm{offset}}=0.077^\circ$ and $\theta_{y,\textrm{offset}}=0.243^\circ$. 
The final inclination angles used in the experiment (Fig. \ref{fig:TiltVariation}) were corrected as $\theta_x=\theta_{x,\textrm{measured}}-\theta_{x,\textrm{offset}}$.
The inclination sensor was positioned inside and outside the vacuum chamber to align the platform's absolute inclination with the external supporting platform.
After adding the magnetic array, the best-fit center of minimal damping was found at ($\theta_x=-0.544^\circ$, $\theta_y=-0.383^\circ$), indicating a small residual tilt.

{
\subsection*{Levitation height measurement}
}
The levitation height $h_{{\rm{V}}}$, defined as the distance between the top surface of the magnet array and the bottom surface of the PG disk, was measured to inform numerical estimations of gas damping in the continuum regime and eddy drag.
A Canon EOS 6D Mark II digital camera with a Sigma 150-600mm F5-6.3 DG OS HSM lens was used for imaging.
To minimize perspective distortion, the camera was positioned approximately 2 m from the setup.
The captured image, along with a length scale, can be found in Supplementary Fig. 1.
A rectangular region within the yellow frame was selected for gray-value analysis using ImageJ. The horizontally averaged gray-value profile is shown as a blue plot.
By leveraging the contrast between the magnet and the levitated disk, the levitation height was estimated to be $h_{{\textrm{V}}}\sim 0.82\:\rm{mm}$.

{
\subsection*{Mechanical oscillation frequency measurement}
}
An active isolation table [TS-300] was used to drive vertical and lateral oscillations. 
The levitation setup was placed on the driving table, and the driving frequencies were swept. 
A dielectric mirror was attached to the disk's center, and displacement was measured using a laser interferometer [Picoscale].
From these measurements, the vertical and lateral mechanical oscillation frequencies of the levitated disk were determined to be $\omega_{{\textrm{V}}}/2\pi={18.9}$ Hz and $\omega_{{\textrm{L}}}/2\pi=\{5.5, 5.9\}$ Hz, respectively. 
While a perfectly symmetric trap would yield identical translational resonance frequencies along $x$ and $y$, two distinct values were observed.

To validate the lateral oscillation frequency, $\omega_{{\textrm{L}}}/2\pi$ was independently measured by exciting lateral motion and analyzing the subsequent ring-down of the disk's COM (Supplementary Fig. 2).
Overhead video recordings at 120 frames per second captured the disk’s motion, with a white dot marking its center. The time dynamics were extracted using Tracker software \cite{Brown2023Tracker6.1.5} and fitted to a decaying sine function, yielding an estimated $\omega_{{\textrm{L}}} /2\pi= 6.0$ Hz.

{
\subsection*{Tilt-horizontal displacement relation}
}
As the setup is inclined by an angle ${\it{\Delta}}\theta$, lateral gravitational forces displace the COM of the PG disk from the magnet's symmetry axis by a distance $d({\it{\Delta}}\theta)$.
In the main text, we estimated for small inclinations that $d({\it{\Delta}}\theta)=g{\it{\Delta}}\theta /\omega_{{\textrm{L}}}^2$.
To verify this relation, we varied ${\it{\Delta}}\theta$ and measured $d$ for a non-rotating PG plate.
The levitation setup was mounted on a breadboard in ambient conditions, and the platform's inclination about the $x-$axis was adjusted using threaded screws.
A Dino-lite Edge 3.0 USB microscope captured images of the disk and an underlying $0.5$ mm grid reticule fixed to the magnet's top surface, serving as a coordinate reference.
Images were taken at inclination angles ranging from 0$^\circ$ to 5$^\circ$, and the displacement $d({\it{\Delta}}\theta)$ was extracted using Tracker software \cite{Brown2023Tracker6.1.5}.
The results, shown in 
{Supplementary Fig. 3}, align well with the theoretical small-angle approximation in Eq. \ref{eq:Tilt_LateralDisplacement_Relation} and support the tilt angle simulation in Fig. \ref{fig:TiltVariation}d, e.

{
\subsection*{COMSOL simulation of rotational eddy damping}
}
The rotational eddy damping rate $\gamma$ as a function of the COM displacement $d$ from the magnet’s axis of symmetry was estimated using a FEM COMSOL simulation.
The model used the same magnet and PG disk dimensions as the experimental setup (Table \ref{table:parameter_values}), while the magnetic and electrical properties (Table \ref{table:simulation_parameters}) were inferred by fitting simulation results to the levitation height and mechanical resonance frequencies.
The eddy-induced damping torque density, generated by the Lorentz force $F_{{\textrm{L}}}$, is given by
\begin{equation}
    \tau_z(x,y,z)=x\times F_{{{\textrm{L}}},y}-y\times F_{{{\textrm{L}}},x}\;\;,
\end{equation}
where $(x,y)$ are coordinates relative to the COM of the PG disk.
{Supplementary Figures 4a, b} show the eddy current distribution and torque density for a disk rotating at $\omega=$ {$\rm{rad}\cdot s^{-1}$} with a small offset $d$.
The total resistive torque $\mathcal{T}$ is obtained by integrating the torque density over the disk volume:
\begin{equation}
    \mathcal{T}=\int_{V_{\mathrm{disk}}}\tau_z \,dV.
\end{equation}
To simulate small tilt angles, we offset the disk’s COM relative to the axisymmetric trap and computed the resulting eddy-damping torque. 
Figure \ref{fig:TiltVariation}e shows the damping rate $\gamma(d)$ extracted from COMSOL simulations. 
As expected, $\gamma$ decreases with decreasing $d$, following a near-perfect power-law dependence for $d>0.05\:\rm{mm}$ (Eq. \ref{eq:power_law_fit}), strongly indicating that $\gamma(d=0)=0$ in an ideally axisymmetric system.
For $d<0.05\:\rm{mm}$, numerical errors (e.g., floating-point precision, finite mesh size, solver accuracy limits) cause $\gamma$ to plateau rather than vanish. 
{Supplementary Figure 4c} illustrates the FEM-generated mesh, which is not perfectly axisymmetric about the trap center, contributing to residual damping at small $d$. 
This simulation does not include gas damping. 
In practice, any inhomogeneity in the graphite disk or magnets, slight inclinations, or other imperfections will break perfect axial symmetry and introduce small eddy-induced torque damping.

\begin{figure*}[tb]
\centering
\includegraphics[width=\textwidth]{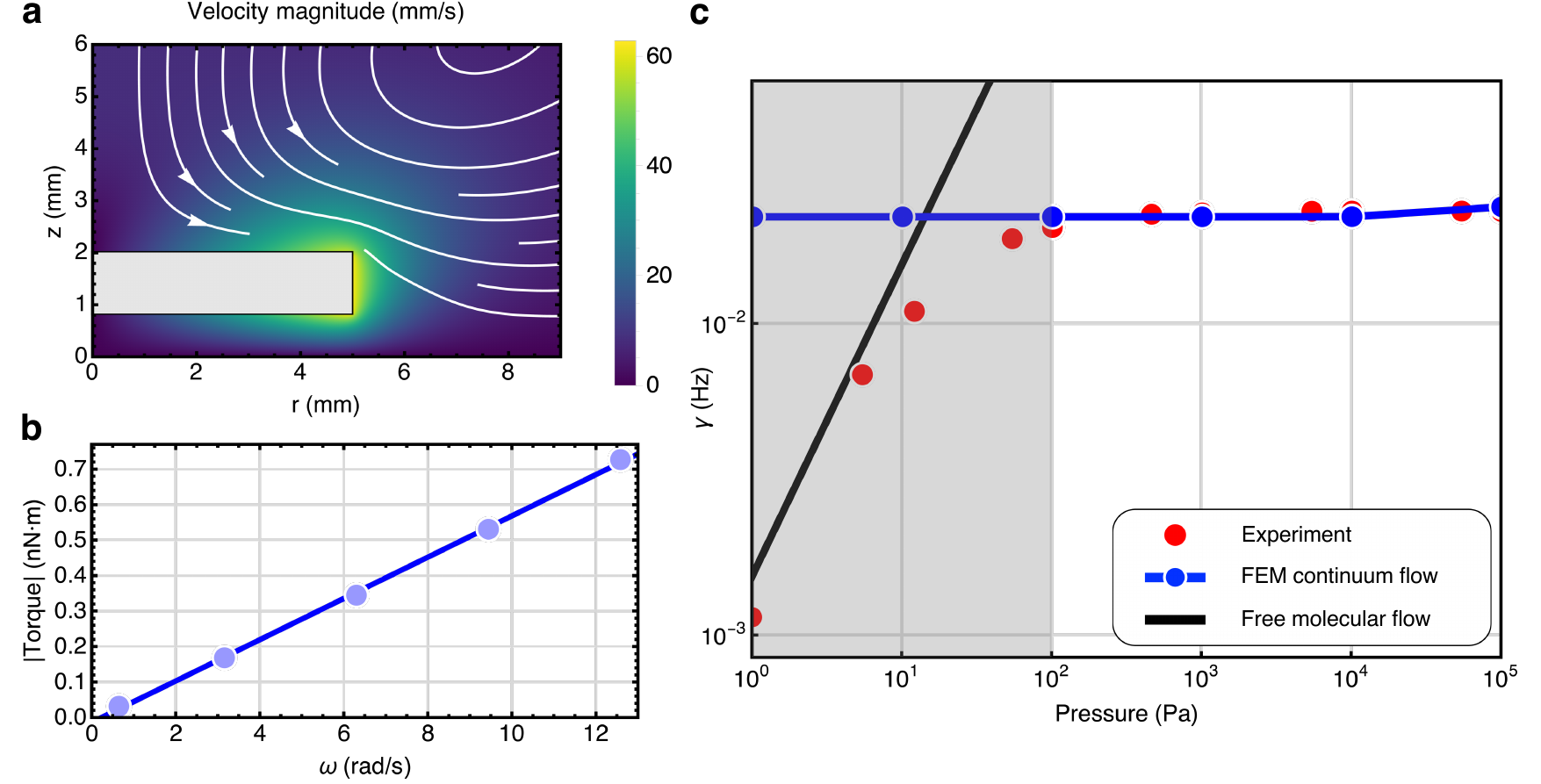}
\caption{
    Simulation of rotational damping in the continuum flow regime (high pressure). 
    A finite element method (FEM) COMSOL fluid dynamic model is developed to predict rotational damping in the regime where the continuum approximation for the gas dynamics holds.
    \textbf{a}: Cross-sectional view of the simulated laminar flow around a rotating disk (gray rectangle) with perfect coaxial symmetry. 
    The disk rotates about its vertical axis $r=0$ at $\omega= 2\pi\times2$ {$\rm{rad}\cdot s^{-1}$} within a cylindrical gas-filled tank in the continuum regime.
    Sliding wall boundary conditions are applied to the disk surfaces.
    Contours and arrows indicate axial flow, while the colormap represents the magnitude of total (axial and azimuthal) flow.
    \textbf{b}: Dependence of the angular damping torque $|\mathcal{T}|$
    on the disk for various angular velocities $\omega$ (light blue circles) at $P\sim 10^5$ Pa, along with the best-fit line (blue).
    The slope of this line corresponds to ${\it{\Gamma}}(P)=\gamma(P)I$.
    \textbf{c}: Comparison of experimental data (red) and FEM simulation results (blue) at high pressure, with {\it no fitted parameters}. 
    {The associated error for the simulation results, obtained from the error in the linear fit, is smaller than the marker size.}
    The simulation aligns well with experimental data at high pressures (unshaded region) but deviates for $P\le 10^2$ Pa (shaded region), where the continuum approximation begins to break down. 
    The solid black line represents the theoretical estimate for $\gamma(P)$ in the molecular flow regime.
    }
\label{fig:comsol_gas_friction}
\end{figure*}

\begin{figure*}[htbp]
\centering
\includegraphics[width=\textwidth]{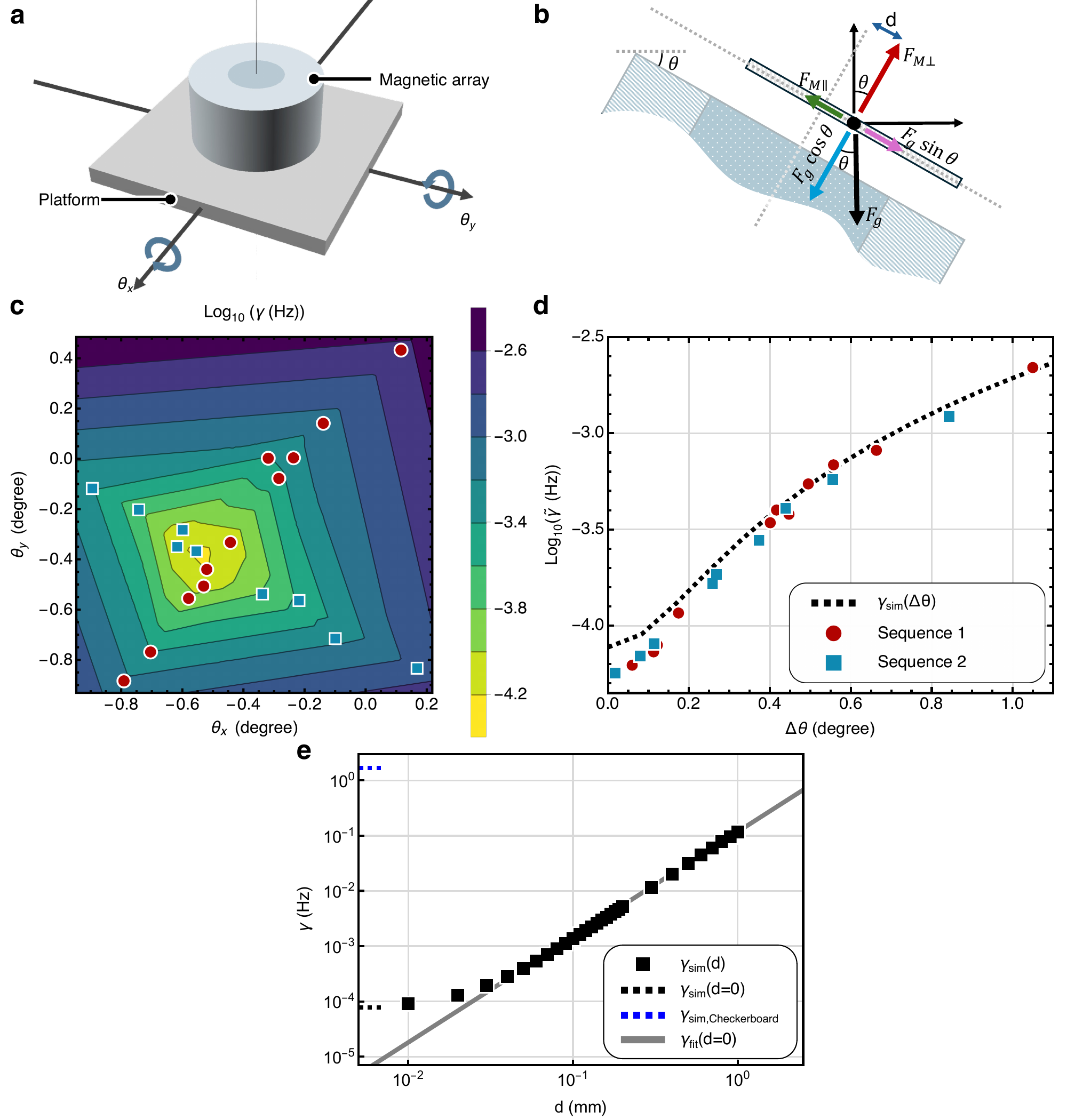}
\caption{
    Studying the dependence of rotational damping rate $\gamma$ on the inclination of the plate to the horizontal: $\gamma(\theta_x,\theta_y)$. 
    As the setup tilts, the center of mass (COM) of the PG disk shifts off axial symmetry, increasing eddy damping.
    \textbf{a}: 
    Tilting of the supporting platform by angles $(\theta_x, \theta_y)$, about the $(x,y)$ axes.
    \textbf{b}: 
    Force diagram of the levitated disk's COM. 
    The PG disk is trapped by a balance of perpendicular forces: $F_{\rm M\perp}=F_{{\textrm{g}}}\cos\theta$, and lateral forces: $F_{\rm M \parallel}\sim m\omega_{{\textrm{L}}}^2\,d=F_{{\textrm{g}}}\sin\theta$,  where $\omega_{{\textrm{L}}}$ is the lateral oscillation frequency of the PG plate, and $d$ is the lateral COM displacement from the axis of symmetry of the magnets, which increases with $\theta$.
    \textbf{c}: 
    Measured dependence of $\gamma(\theta_x,\theta_y)$.
    Red circles and blue squares indicate two measurement sequences, showing $\min \gamma(\theta_x,\theta_y)$ at $(\theta_x,\theta_y)=(\theta_x^0, \theta_y^0)\ne (0,0)$ due to a small initial misalignment.
    {The associated error (SE) is smaller than the marker size.}
    \textbf{d}: 
    Radial fit to $\gamma(\delta\theta_x+\theta_x^0,\delta\theta_y+\theta_y^0)\equiv \tilde{\gamma}({\it{\Delta}} \theta)$, where ${\it{\Delta}}\theta^2=\delta\theta_x^2+\delta\theta_y^2$, yielding a 1D collapsed plot of the 2D data from {\bf c}.
    {The associated vertical and horizontal errors for the measurements, obtained from the error in the linear fit and SE, are smaller than the marker size.}
    The dashed curve is a FEM-COMSOL estimate assuming only eddy damping, matching experiments for ${\it{\Delta}} \theta > 0.1^\circ$ but breaking down for ${\it{\Delta}} \theta < 0.1^\circ$ (see {\bf e}).
    \textbf{e}: 
    Numerical FEM study of $\gamma(d)$. 
    From symmetry arguments, we expect $\gamma(d\rightarrow 0)\rightarrow 0$, but due to the non-axially symmetric nature of the FEM mesh and numerical errors, the simulation yields $\gamma(d\rightarrow 0)\ne 0$. 
    For $d> 0.05$ mm, simulations show a near-perfect power-law dependence (Eq. \ref{eq:power_law_fit}), 
    supporting $\gamma(d=0)=0$ for perfect axial symmetry.
    }
\label{fig:TiltVariation}
\end{figure*}

{
\subsection*{COMSOL modeling of gas damping at atmospheric pressure}
}
COMSOL simulations were performed to model gas friction damping on the levitated disk in the continuum regimes corresponding to near-atmospheric pressure.
To simplify the simulation, we modeled a rotating airflow around a stationary disk near a surface and evaluated the gas-induced torque acting on the disk.
Using COMSOL's laminar flow module, we solved the Navier-Stokes equation with a sliding-wall boundary condition on the disk surfaces, specifying an azimuthal fluid flow of $r\omega$, where $r$ is the radial distance from the axis of symmetry.
A no-slip boundary was applied at $z=0\:\rm{mm}$, representing the magnetic array's top surface, while symmetric boundary conditions were used for the top and right boundaries, representing free surfaces.
The levitation height was set to $h_{{\textrm{V}}}= 0.82\:\rm{mm}$, as measured in Supplementary Fig. 1.
The gas density $\rho_{\rm{gas}}$ was determined using the ideal gas law for dry air, $\rho_{\rm{gas}} = P/(R_{{\textrm{s}}} T)$, while the dynamic viscosity $\mu$ was treated as pressure-independent and kept constant.
Figure \ref{fig:comsol_gas_friction}a shows the computed airflow at atmospheric pressure. 
The damping torque $\mathcal{T}$ was calculated as the total shear stress on the disk's surface, where the shear stress is given by the shear rate times $\mu$.
The torque was evaluated over rotational speeds  $\omega\in 2 \pi \times\{0.1, 2.0\}$ {$\rm{rad}\cdot s^{-1}$}.
A linear fit to $(\omega,\mathcal{T})$ yielded the damping coefficient in the continuum regime, $\mathcal{T}=|{\it{\Gamma}}_{{\textrm{c}}}|\omega$, as shown in Fig. \ref{fig:comsol_gas_friction}b. 
From this, the damping rate was derived as $\gamma_{{{\textrm{c}}}}={\it{\Gamma}}_{{{\textrm{c}}}}/I$.
Simulated damping rate estimates $\gamma_{{\textrm{c}}}$ were obtained over a pressure range of 1 to $10^5$ Pa and compared with the experimental results (Fig. \ref{fig:comsol_gas_friction}c).
The simulation agrees well with the experimental observations at high pressures ($P>10^2$ Pa) but deviates at lower pressures, where the continuum assumption begins to break down.

{
\section*{Data availability}
}
All relevant data are available in the article and Supplementary Information or are available from the corresponding author upon reasonable request.

{
\section*{Acknowledgments}
}
This work was supported by the Okinawa Institute of Science and Technology (OIST), Japan. 

We gratefully acknowledge the Engineering Section and the Scientific Computing and Data Analysis Section at OIST for their help and support, and C.G. George for assistance with photography.

{
\section*{Author contributions} 
}
D.K., S.T., and J.T. designed the study. D.K., S.T., B.C., C.S.J., and J.D. conducted the experimental work. D.K. and S.T. performed the simulations. D.K., S.T., and J.T. analyzed the results and prepared the manuscript. All authors revised the manuscript prior to submission.

{
\section*{Competing interests}
}
The authors declare no competing interests.



\begin{figure*}[tb]
\centering
\includegraphics[width=0.5\textwidth]{Fig5.png}
\caption{
    Comparison between the simulated magnetic field generated by a typical axially symmetric magnetic array and the analytical example.
    The simulated magnetic field {$\mathbf{B}_{\rm{ex}}$} is calculated with an exact analytical expression \cite{CACIAGLI2018423} with the parameters listed in the Table. \ref{table:parameter_values} and \ref{table:simulation_parameters}.
    The analytical resembling magnetic field, { {$\mathbf{B}_{\rm{res}}$} $=B_r(r) \:\hat{r} +B_z(r,z)\:\hat{z}$}, is built from Eq. \ref{eq:analytical_B_field_example}, with $\alpha=\pi/6$, $\beta=0.1$, and $z_0=-4$.
    }
\label{fig:B_field_comparison_analytical}
\end{figure*}


\begin{table}[h]
    \centering
    \renewcommand{\arraystretch}{1.2} 
    \caption{{Parameters used in the rotational eddy damping simulations. The table lists the magnetizations of the N52 cylindrical and N40 ring magnets (manufacturer-specified values, marked with *), as well as the anisotropic magnetic susceptibilities and electrical conductivities of pyrolytic graphite (PG). Magnetic susceptibilities are dimensionless, while conductivities are given in $\rm{S\cdot m^{-1}}$. These parameters were used as input for the finite-element simulations of rotational damping.}}
    \begin{tabular}{|m{6cm}|m{1cm}|m{1.5cm}|m{1cm} |} 
        \hline\hline
        \textbf{Simulation parameters}& \textbf{Symbol}& \textbf{Value}& 
        \textbf{Units}  \\ 
        \hline
        \hline
        N52 cylindrical magnet magnetization*  & & 1480 &  mT  \\ 
        N40 ring magnet magnetization* & & 1300 &  mT  \\ 
        PG horizontal volume magnetic susceptibility &$\chi_\parallel$ & $85\times 10^{-6}$ &  1  \\ 
        PG vertical volume magnetic susceptibility& $\chi_\perp$ & $530\times 10^{-6}$ &  1  \\ 
        PG horizontal electrical conductivity & $S_\parallel$ & 130,000 & {$\rm{S}\cdot m^{-1}$}  \\ 
        PG vertical electrical conductivity &$S_\perp$& 200 & {$\rm{S}\cdot m^{-1}$}  \\ 
        \hline       
        \hline
    \end{tabular}
    \label{table:simulation_parameters}
\end{table}

\begin{table}[h]
    \centering
    \renewcommand{\arraystretch}{1.2} 
    \caption{{Geometric, material, and measured parameters used in this study. The table lists the dimensions of the magnets and PG disk, the measured mass and levitation properties, the trap frequencies, and the ambient air density and viscosity values used in COMSOL gas-torque simulations.}}
    \begin{tabular}{|m{6cm}|m{1.2cm}|m{2cm}|m{1cm} |} 
        \hline\hline
        \textbf{Description of Quantity} & \textbf{Symbol} & \textbf{Value}& \textbf{Units}  \\ 
        \hline\hline
        N40 ring magnet outer diameter  &  & 19 ($\pm$0.1) &  mm  \\ 
        N40 ring magnet inner diameter  &  & 8.1 ($\pm$0.1) &  mm  \\ 
        N40 ring magnet height & & $4\:(\pm0.1)\times 5$ & mm\\
        N52 cylindrical magnet diameter  &  & 8 ($\pm$0.1) &  mm  \\ 
        N52 cylindrical magnet height & & $10\:(\pm0.1)\times 2$ & mm\\
        PG disk diameter & $D$ & 10.02 ($\pm$0.01) & mm\\
        PG disk thickness & $H$ & 1.12 ($\pm$0.01) & mm\\
        PG mass & $M$ & 191 ($\pm$1)&  mg \\
        Measured levitation height & $h_{{\textrm{V}}}$ & 0.82 & mm\\
        Measured vertical trap frequency&   $\omega_{{\textrm{V}}}/2\pi$& 18.9 & Hz\\
        Measured lateral trap frequency&   $\omega_{{\textrm{L}}}/2\pi$& 6.0 & Hz\\
        The density of air at ambient for COMSOL gas torque simulation & $\rho_{\rm{gas}}$ & 1.204& {$\rm{kg}\cdot m^{-3}$}\\
        The viscosity of air at ambient for COMSOL gas torque simulation & $\mu$ & $1.81\times10^{-5}$ & Pa$\cdot$s\\         
        \hline       
        \hline
    \end{tabular}
    
    \label{table:parameter_values}
\end{table}

\end{document}


\title[Article Title]{Supplementary information for: A magnetically levitated conducting  rotor with ultra-low rotational damping circumventing eddy loss}


\author*[1]{\fnm{Daehee} \sur{Kim}}\email{daehee.kim@oist.jp}
\author*[1]{\fnm{Shilu} \sur{Tian}}\email{shilu.tian@oist.jp}

\author[1]{\fnm{Breno} \sur{Calderoni}}
\author[1]{\fnm{Cristina Sastre} \sur{Jachimska}}
\author[2]{\fnm{James} \sur{Downes}}
\author*[1]{\fnm{Jason} \sur{Twamley}}\email{jason.twamley@oist.jp}

\affil*[1]{\orgdiv{Quantum Machines Unit}, \orgname{Okinawa Institute of Science and Technology Graduate University}, \orgaddress{\city{Onna}, \postcode{904-0495}, \state{Okinawa}, \country{Japan}}}

\affil[2]{\orgdiv{School of Mathematical and Physical Sciences}, \orgname{Macquarie University}, \state{NSW} \postcode{2109}, \country{Australia}}



\maketitle

\section*{Supplementary Note 1: FEM modeling}
\begin{figure*}[tb]
\centering
\includegraphics[width=\textwidth]{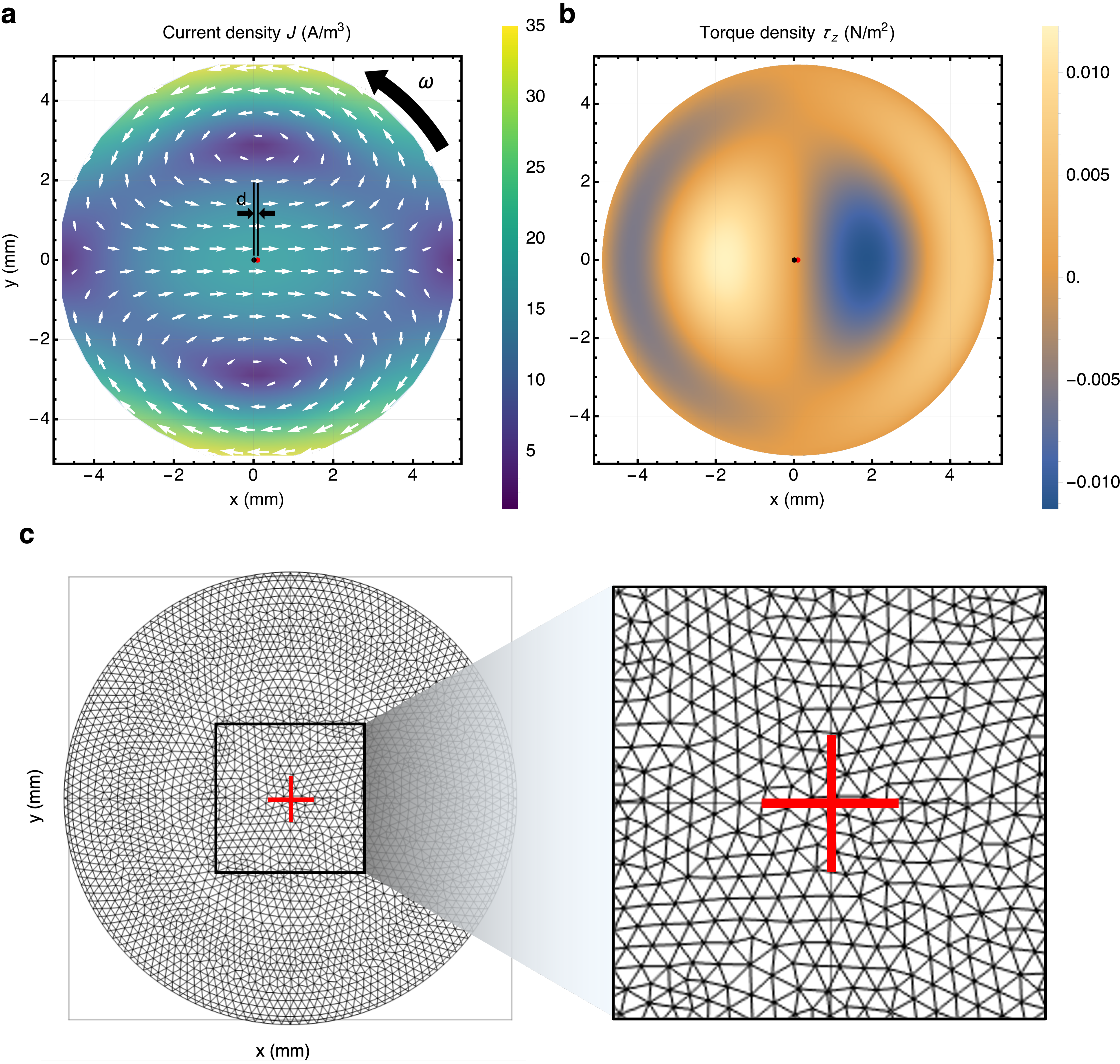}
\caption{
    The FEM COMSOL simulation of eddy current distribution and torque density in an offset rotating levitated PG disk.
    The disk's center of mass (COM, red dot) is laterally offset by $d= \rm 0.1 \: mm$ from the magnet's axis of symmetry (black dot), and the disk is set to have an angular velocity $\omega= 15 \rm\: rad/s$.
    \textbf{a}: Eddy currents flowing on the \(x\)-\(y\) top surface of the disk.
    Colors indicate current density magnitude \(J\), while arrows show the current flow direction in the lab frame.
    \textbf{b}: \(z\)-component of the torque $\tau_z$ on the disk's top surface along the center, generated by eddy currents.
    \textbf{c}:
    Non-axisymmetric top-surface mesh.
    Despite the disk being aligned with the magnet’s axis, the asymmetric mesh introduces residual torque, leading to artificial eddy damping in the simulation.
    }
\label{fig:comsol_eddy_current}
\end{figure*}

As mentioned in the main manuscript, the FEM simulation requires parameter values associated with our setup to model the eddy damping of the off-axis rotations.
We measured the levitation height, vertical and lateral oscillation frequencies (refer to Methods), and these values are shown in Table 2. Using these we deduced the appropriate values of $(\chi_\parallel,\chi_\perp)$ (see Table 1) for our graphite material and used FEM to model our setup to yield simulation values: $h_{\textrm{V},\mathrm{sim}} = 0.83\,\mathrm{mm}$, $\omega_{{\textrm{V}},\mathrm{sim}}/2\pi = 19.0\,\mathrm{Hz}$, $\omega_{{\textrm{L}},\mathrm{sim}}/2\pi=5.8\,\mathrm{Hz}$, which agree very well with the measured values.
The estimated values of the susceptibilities, together with fitted values for $(S_\parallel, S_\perp)$ (see Table 1), were then used to simulate the rotational eddy damping, $\gamma_\mathrm{sim}(d)$.
\\
\indent
Supplementary Figure \ref{fig:comsol_eddy_current}c shows an example of the generated FEM mesh at the top surface of the disk. 
Although we used an extremely fine mesh to resolve the disk, the mesh is clearly not axisymmetric about the center of the disk, and this breaking of axial symmetry of the  FEM mesh will lead to effective non-vanishing eddy currents.
To clearly see the effect of eddy currents on the rotational damping, we arbitrarily shift the disk $d=0.1$ mm off center in the simulation and plot the current density $\textbf{J}$ and torque density $\tau_z$ distribution generated on the top surface of the disk, as shown in 
Supplementary Fig. \ref{fig:comsol_eddy_current}a and b.
The center of the trap and the center of the disk are marked with black and red dots, respectively. 
Even a slight shift (e.g., $d=0.1$ mm) of the disk can break the axial symmetry and create circulating eddy currents, which generate a damping torque on the rotational motion. 
These simulation results imply that any horizontal inhomogeneity in the graphite disk and magnets leads to eddy damping on the disk's rotation, even if the geometric center of the disk is co-axial with the geometric center of the magnets. 

\begin{figure*}[tb]
\centering
\includegraphics[width=\textwidth]{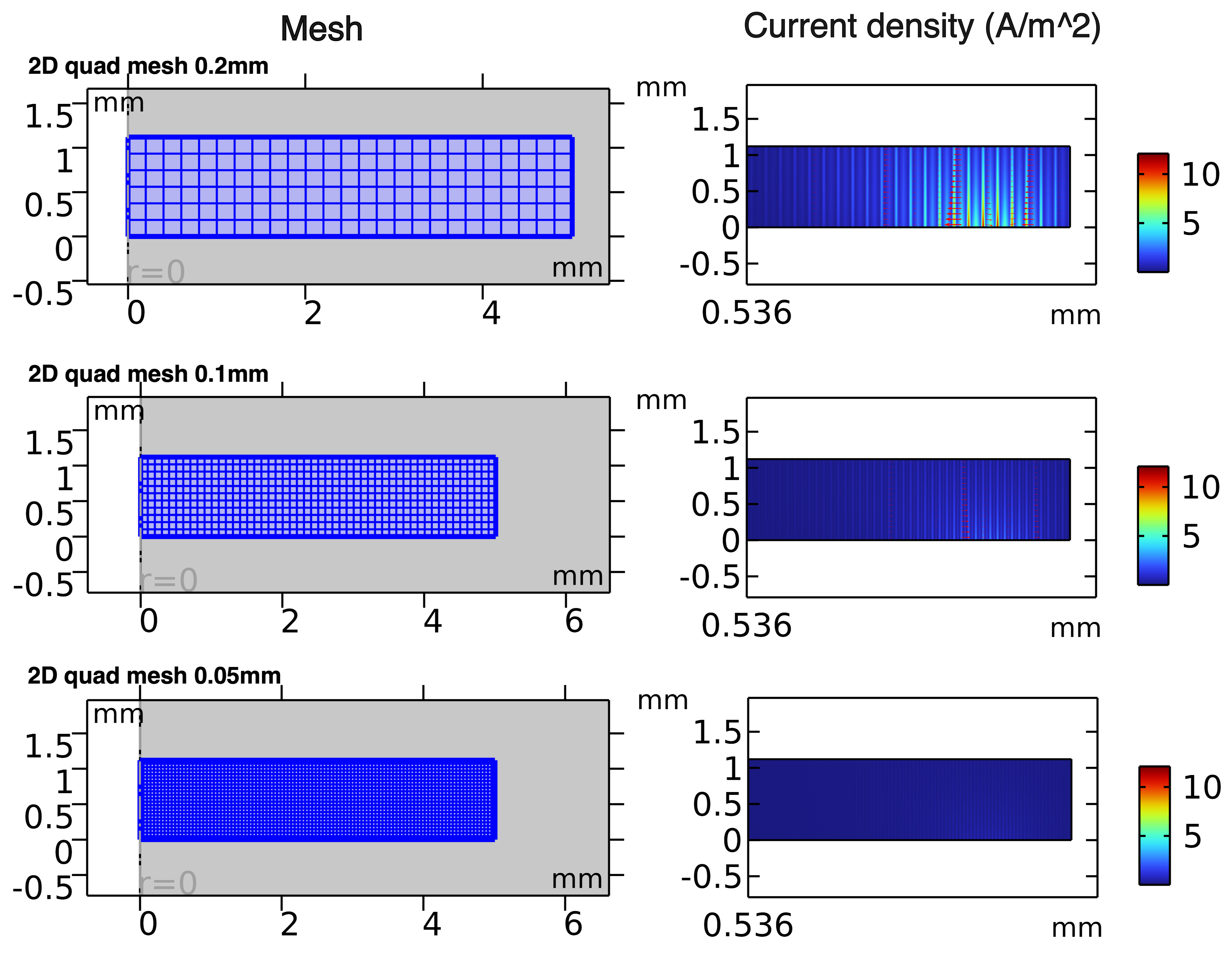}
\caption{
    2D axisymmetric FEM simulation of the eddy currents using \textcolor{red}{a} quad mesh. The left column is the mesh with different sizes\textcolor{red}{,} and the right column is the spurious current density. The current distribution is controlled by the mesh. With the mesh getting finer, the spurious currents get lower.  
    }
\label{fig:2D_Quad}
\end{figure*}

\begin{figure*}[tb]
\centering
\includegraphics[width=\textwidth]{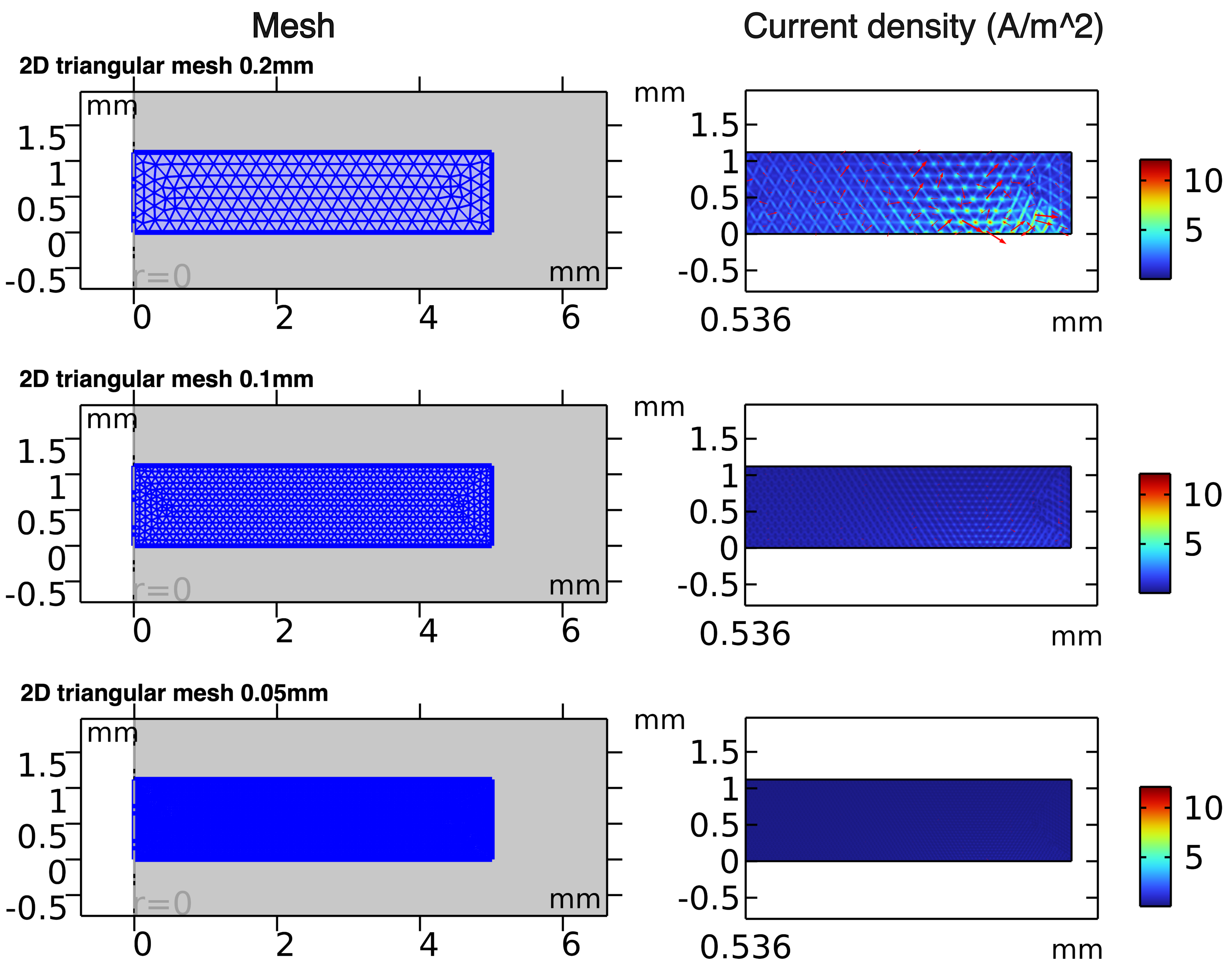}
\caption{
    2D axisymmetric FEM simulation of the eddy currents using \textcolor{red}{a} triangular mesh. The left column is the mesh with different sizes\textcolor{red}{,} and the right column is the spurious current density. It shows the same mesh-dependent effect. The current distribution is controlled by the mesh. With the mesh getting finer, the spurious currents get lower.  
    }
\label{fig:2D_Triangular}
\end{figure*}

To reduce the numerical error caused by the non-symmetric mesh, we then build a 2D-axisymmetric FEM model. We get a lower damping rate ($\gamma \sim 4\times 10^{-6} \:\rm Hz$) from the 2D simulation. 
Supplementary Figures \ref{fig:2D_Quad} and \ref{fig:2D_Triangular} show that the currents and damping rate are decreased with reducing the mesh size, and the spurious currents are controlled by the mesh. This confirms that the residual damping is caused by the spurious currents in the numerical artifacts.

\begin{table}[ht]
\centering
\caption{Damping rates as a function of mesh size for quad, triangular, and 3D meshes.}
\label{tab:mesh-damping}
\begin{tabular}{llll}
\toprule
Mesh size (mm) & Damping 2D quad mesh      & Damping 2D triangular mesh   & Damping 3D model     \\
\midrule
0.20           & $7.91\times10^{-6}$      & $6.50\times10^{-6}$         & $7.72\times10^{-5}$ \\
0.10           & $4.24\times10^{-6}$      & $4.16\times10^{-6}$         &                      \\
0.05           & $4.01\times10^{-6}$      & $4.00\times10^{-6}$         &                      \\
\bottomrule
\end{tabular}
\end{table}

\begin{figure}
\includegraphics[width=\columnwidth]{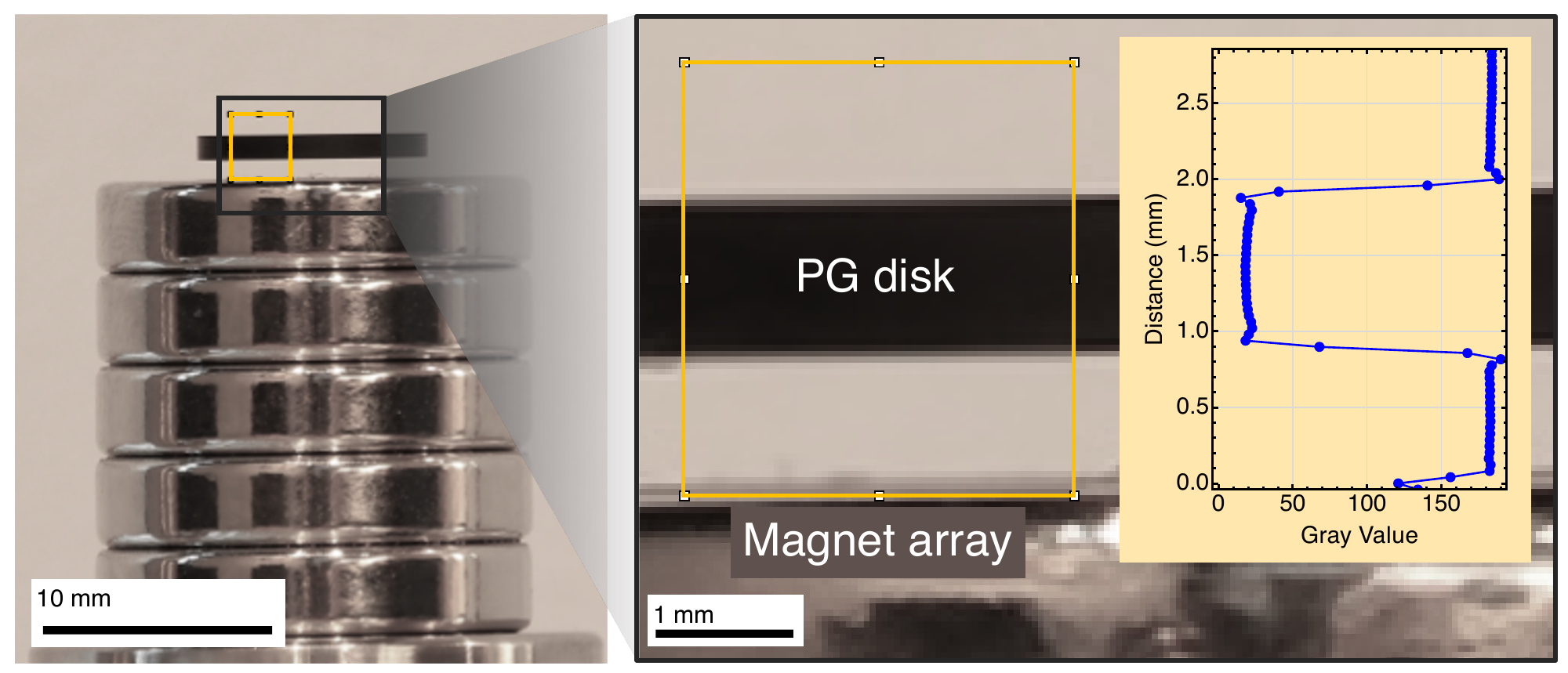}
\caption{Levitation height measurement.
A high-resolution photograph of the experimental setup.
A photograph of the setup, including a length scale, was taken for precise measurements.
The inset provides a magnified view of the top surface of the magnet, the levitating disk, and the gap between them, representing the levitation height.
A yellow-outlined rectangular region was selected to analyze the horizontally averaged gray values, shown in the blue plot on the right.
The estimated levitation height for this measurement is $h_{{\textrm{V}}}=0.82\:\rm{mm}$, representing the distance between the top surface of the magnets and the bottom surface of the PG disk.
This value serves as a validation check for our FEM simulation.
}\label{fig:LevitationHeight}
\end{figure}

\begin{figure}
\includegraphics[width=\columnwidth]{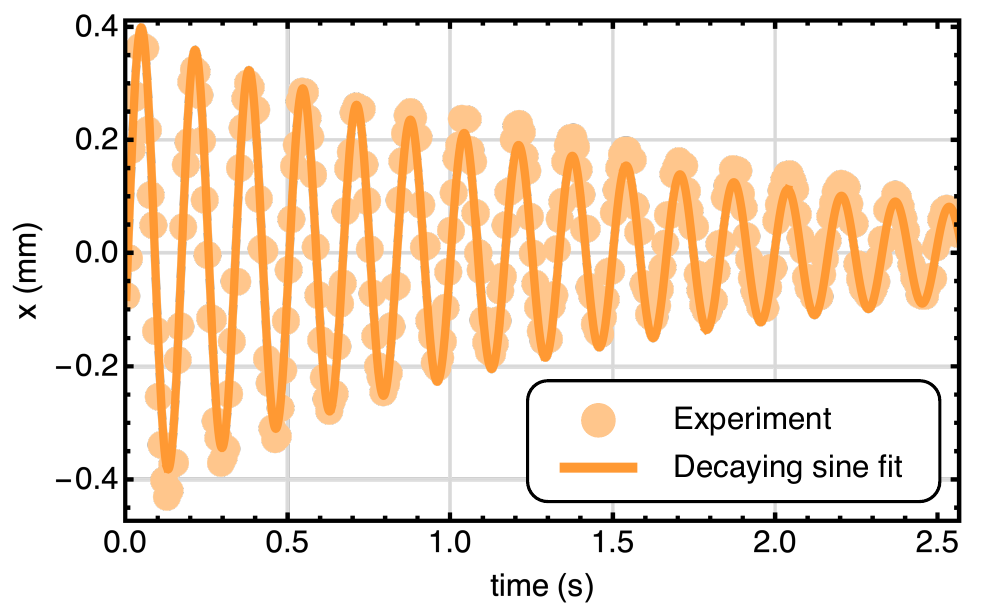}
\caption{
Lateral trap frequency measurement.
Lateral oscillations of the levitated PG plate are excited, and the ring-down of the motion is analyzed to estimate the lateral trapping frequency. 
Due to significant eddy damping, these translational oscillations decay rapidly.
The $ x-$coordinate of the levitating disk's center is tracked by recording a video and analyzing the motion using Tracker software [https://physlets.org/tracker/]. 
The measured positions are fitted to a damped sine wave, yielding a lateral oscillation frequency of $\omega_{{{\textrm{L}}}}/2\pi = 6.0$ Hz. 
This frequency is then used to estimate the horizontal displacement $d$ of the levitated plate when the setup is inclined, as applied in the FEM simulation in Fig. 4.
}
\label{fig:LateralOscillation}
\end{figure}

\begin{figure}
\includegraphics[width=\columnwidth]{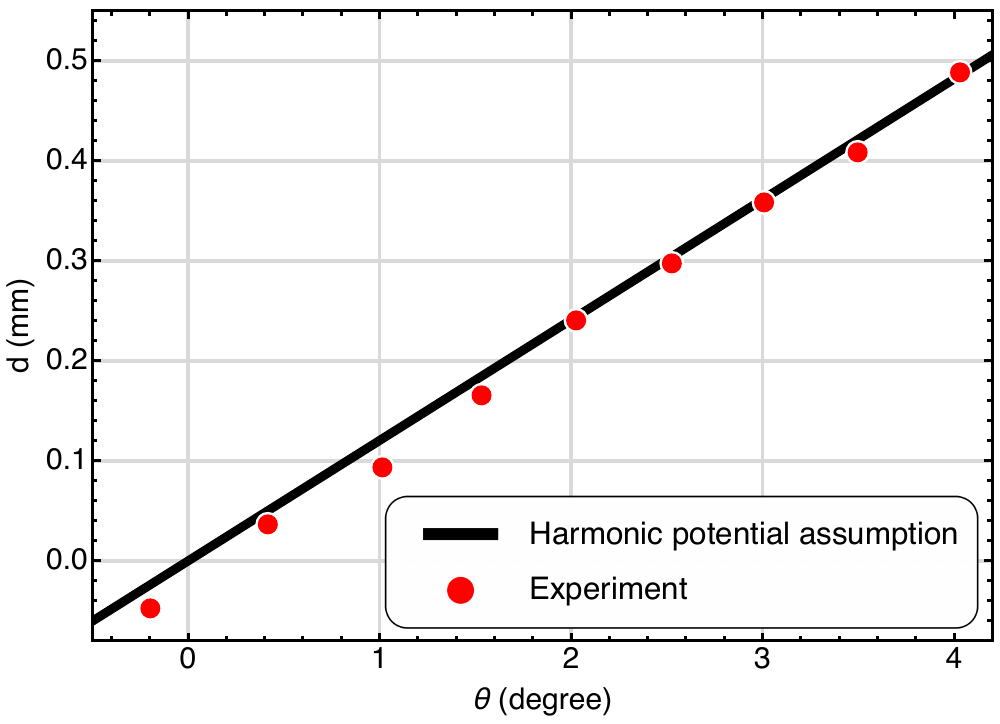}
\caption{
Tilt angle to horizontal displacement relationship.
As the setup is inclined, the levitated disk shifts due to gravity, causing it to rotate about an axis slightly offset from the magnet's central axis by a distance $d$. 
This displacement increases rotational eddy damping. 
The displacement $d$ can be estimated using the known lateral trap frequency.
As a verification, $d$ is also measured as a function of the tilt angle $\theta$ (red dots) using the Tracker software.
The theoretical prediction, based on the lateral harmonic trap frequency of $\omega_{{\textrm{L}}}/2\pi = 6.0$ Hz (solid black line), shows good agreement with the measured values.
}\label{fig:TiltLateralDisplacementRelation}
\end{figure}

\vspace{.5cm}
 
\noindent

\vspace{.5cm}
 
\noindent

\section*{{Supplementary Note 2:} Comparison with rotation damping rates in the literature}
We can compare the minimum damping rate we observe in our axisymmetric diamagnetic rotor with other levitated rotational studies reported in the literature. These include a diamagnetically levitated graphite-acrylic ring (Waldron, 1966 \cite{Waldron1966DiamagneticGraphite}), a rotating steel sphere which is magnetically levitated with active stabilization (Fremerey, 1971 \cite{Fremerey1971ApparatusSuspension}), an optically levitated 10 $\mu$m silica sphere (Monteiro, 2018 \cite{Monteiro2018OpticalVacuum}), a 100nm silica sphere (Ahn, 2020 \cite{Ahn2020UltrasensitiveNanorotor}), a silica dumbell made from silica spheres 140nm diameter (Ju, 2023 \cite{Ju2023Near-FieldNanodumbbell}), a 15 mm diameter spherical NdFeB magnet levitating above a superconductor (Druge, 2014 \cite{Druge2014DampingSuperconductor}). We note that the quantized energy of a quantum rotor is $E_l\sim \hbar^2 l(l+1)/2I$, and to see the quantization of this energy, one would like $\Delta E_l=E_{l+1}-E_l$ to be large, requiring a small moment of inertia $I$. However, the rotational damping rate, given a fixed gas pressure, decreases with increasing inertia $I$. We can see this by studying the formula for the rotational damping rates for a sphere of diameter $D$, density $\rho_{{\textrm{s}}}$, in a gas at pressure $p$, either in the free molecular flow regime or continuum regime:
\begin{equation}
    \gamma_{\rm{fm}}\approx \frac{p}{\rho_{{\textrm{s}}} D}\sqrt{\frac{2m_{0}}{\pi k_{{\textrm{B}}} T}}\;\;, \qquad\qquad
\gamma_{{{\textrm{c}}}}\approx\frac{\eta}{\rho_{{\textrm{s}}} D^2}\;\;\label{gammac}
\end{equation}
where 
$\eta$ is the continuum gas viscosity. We list the rotational damping rates $\gamma_{\rm{fm}}$ and $\gamma_{{\textrm{c}}}$, for our work and the above references in 
{Supplementary} Table \ref{table:comparison}.
\begin{table}[ht]
\centering
\caption{Reported rotor damping rates, volumes, and vacuum levels from the literature and our lowest damping rate [row 1]. We indicate whether in the free molecular flow [fp] or continuum [c] regimes.}
\label{tab:damping-mbar}
\begin{tabular}{llll}
\toprule
Reference & Damping Rate $\gamma$ (Hz) & Volume (m$^3$) & Vacuum (mbar) \\
\midrule
Kim (2025)  [fm]        & $5\times10^{-5}$                  & $3\times10^{-7}$             & $4.85\times10^{-7}$ \\
Waldron (1966) [fm]     & $22\times10^{-6}$                 & $2.83\times10^{-6}$          & $3.73\times10^{-5}$ \\
Fremerey (1971) [fm]    & $\sim10^{-9}$                     & $2.25\times10^{-8}$          & below $1.33\times10^{-9}$ \\
Monteiro (2018)  [fm]   & $1.25\times10^{-5}$ (Fig.\,4a)    & $5.2\times10^{-16}$          & $\sim 10^{-7}$           \\
Ahn (2020) [fm]         & $5\times10^{-2}$ (from Fig.\,3b)  & $1.77\times10^{-21}$         & $2.67\times10^{-6}$ \\
Ju (2023)   [fm]        & $5\times10^{-2}$ (from Fig.\,2b)  & $3\times10^{-21}$            & $1.00\times10^{-5}$ \\
Druge (2014) [c]   & $8.3\times10^{-4}$ (Fig.\,3 at 15\,mm) & $1.77\times10^{-6}$     & $1.01\times10^{3}$  \\
\bottomrule
\end{tabular}\label{table:comparison}
\end{table}

\section*{{Supplementary Note 3:} Skin depth calculation}
We investigate whether the skin effect is negligible in our study. When one has an AC current flowing in an electrical conductor, under certain conditions, it will flow in a layer concentrated around the outer surface of the conductor with a transverse layer thickness of $\delta$. So if the body is rotating off-axis, the body will experience a time-dependent magnetic field and this may induce an AC eddy current. 
Skin depth $\delta$ for a conductor is
\begin{equation}
    \delta=\frac{2}{\omega \mu \sigma},
\end{equation}
where $\omega/2\pi$ is the frequency of AC, $\mu$ is permeability of the conductor, and $\sigma$ is the conductivity of the conductor.
Using $\sigma_\parallel=1.3\times 10^5\:\rm{S{{\cdot\textrm{m}^{-1}}}}$ and $\sigma_\perp=2.0\times 10^2\:\rm{S{{\cdot\textrm{m}^{-1}}}}$ from Table 1, we obtain that $\delta_{\parallel}=0.14\;\rm{m}$ and $\delta_{\perp}=3.6\;\rm{m}$.
Therefore, skin effect, or the concentration of any AC eddy currents to the surface, is negligible in our study.

 



\renewcommand{\refname}{{Supplementary references}}
